# Suicide-Gene Transfection of Tumor-tropic Placental Stem Cells employing Ultrasound-Responsive Nanoparticles

Juan L. Paris,[a,b] Paz de la Torre,[c] M. Victoria Cabañas,[a] Miguel Manzano,[a,b] Ana I. Flores*[c] and María Vallet-Regí*[a,b]

[a]Dpto. Química en Ciencias Farmacéuticas, Facultad de Farmacia, UCM, Instituto de Investigación Sanitaria Hospital 12 de Octubre (imas12), 28040-Madrid, Spain. E-mail: vallet@ucm.es

[b]Centro de Investigación Biomédica en Red de Bioingeniería, Biomateriales y Nanomedicina (CIBER-BBN), Spain

[c]Grupo de Medicina Regenerativa, Instituto de Investigación Sanitaria Hospital 12 de Octubre (imas12), Madrid, Spain. E-mail: anaisabel.flores@salud.madrid.org

**Abstract**

A Trojan-horse strategy for cancer therapy employing tumor-tropic mesenchymal stem cells transfected with a non-viral nanovector is here presented. In this sense, ultrasound-responsive mesoporous silica nanoparticles were coated with a polycation (using two different molecular weights), providing them with gene transfection capabilities that were evaluated using two different plasmids. First, the expression of Green Fluorescent Protein was analyzed in Decidua-derived Mesenchymal Stem Cells after incubation with the silica nanoparticles. The most successful nanoparticle was then employed to induce the expression of two suicide genes: cytosine deaminase and uracil phosphoribosyl transferase, which allow the cells to convert a non-toxic pro-drug (5-fluorocytosine) into a toxic drug (5-Fluorouridine monophosphate). The effect of the production of the toxic final product was also evaluated in a cancer cell line (NMU cells) co-cultured with the transfected vehicle cells, Decidua-derived Mesenchymal Stem Cells.

## 1. Introduction

Cell-based cancer therapies have attracted great attention with many different cell types being proposed due to their tumor-tropic behavior [1–5]. Mesenchymal Stem Cells (MSCs) have been extensively evaluated for cancer therapy [5] because of their inherent tumor-suppressing capacities [6,7], as well as by their capacity to be transfected with therapeutic genes [8,9], or by acting as carriers of nanodrugs [3,10].

Regarding therapeutic approaches involving gene transfection, a promising strategy consists in inducing the expression of a suicide gene [11]. Suicide genes can act either directly, encoding a protein that is toxic, or indirectly, encoding an enzyme that can lead to the production of a toxic molecule [11]. Tumor-tropic cells transfected with suicide genes can induce cell death of both, the transfected vehicle cell and the surrounding cancer cells [8,9,12,13], in a phenomenon known as the "bystander effect" [14]. However, most of these approaches have relied on viral vectors as transfection agents, which present several safety concerns regarding their clinical application [8,15,16]. Then, gene transfection using non-viral vectors might be more suitable for developing therapeutic strategies [12].

Likewise, a particularly interesting type of MSCs are Decidua-derived Mesenchymal Stem Cells (DMSCs) obtained from human placenta [17], which have been shown to inhibit tumor progression [6]. We have recently reported the use of DMSCs as vehicles capable of transporting Mesoporous Silica Nanoparticles (NPs) and Ultrasound-responsive NPs (UR-NPs) towards tumors, both *in vitro* [17]and *in vivo* [18]. These UR-NPs can be loaded with doxorubicin (a cytotoxic drug) and then, upon ultrasound exposure, the drug is released-killing surrounding cancer cells [17].

The objective of the present work is to use engineered UR-NPs as non-viral transfection agents of a suicide gene to transform DMSCs into a Trojan-horse for cancer therapy. DMSCs will migrate towards tumor tissue while producing the enzyme which, upon administration of a non-toxic pro-drug, will generate highly toxic species capable of killing the surrounding cancer cells. Moreover, a critical advantage of using a stimulus-responsive nanocarrier, such as UR-NPs, is that it could enable a combined strategy in which both, small therapeutic molecules, like doxorubicin, and a plasmid, are delivered to the vehicle cells. In this combined strategy, DMSCs would uptake nanoparticles carrying both, an anticancer drug and a plasmid encoding suicide genes for further systemic injection. After migration towards tumor tissue, a therapeutic effect will be produced by the conversion of an injected non-toxic pro-drug into toxic species. Afterwards, another therapeutic effect will happen upon application of ultrasound in the area of interest, which would induce the release of the anticancer drug. This approach could drastically improve the therapeutic outcome by exploiting synergisms between both mechanisms of action [19]. The non-toxic nature of the employed pro-drug and the avoidance of premature release of the anticancer drug by UR-NPs would significantly reduce undesired side effects of anticancer treatments.

## 2. Materials and methods

### 2.1 Synthesis and evaluation of polycation-coated UR-NPs

*Synthesis of UR-NPs*: Ultrasound-responsive Mesoporous Silica Nanoparticles were obtained as described elsewhere [20], by grafting the polymeric gate poly-(2-(2-methoxyethoxy)ethyl methacrylate-co-2-tetrahydropyranyl methacrylate), p($MEO_2MA$-co-THPMA), to mesoporous silica nanoparticles. Those MCM-41 type mesoporous silica nanoparticles were prepared following a modified Stöber method, based on the condensation of Tetraethyl orthosilicate under dilute conditions in the presence of Hexadecyltrimethylammonium bromide [20]. The surfactant was removed by ionic exchange with $NH_4NO_3$ and the particles were collected by centrifugation and washed with ethanol [18,20]. The ultrasound-responsive copolymer p($MEO_2MA$-co-THPMA) was obtained by Free Radical Polymerization of $MEO_2MA$ and THPMA monomers (90:10 molar ratio in the final polymer)[20]. The copolymer was then grafted on the surface of Mesoporous Silica Nanoparticles by conjugating it with 3-Aminopropyltriethoxysilane (APTES) through dicyclohexylcarbodiimide/*N*-Hydroxysuccinimide (DCC/NHS) chemistry, and then through the condensation of the APTES moieties on the silica surface [17,20].

*Polyethylenimine (PEI) coating of UR-NPs*: UR-NPs were coated with linear PEI of two molecular weights (1.8 kDa or 5 kDa), obtaining UR-NPs@1.8PEI and UR-NPs@5PEI, respectively. UR-NPs@PEI were obtained by dissolving 5 mg of PEI (1.8 or 5 kDa) in 0.5mL of deionized (DI) water with 3 µL of acetic acid (glacial). The PEI solution was then diluted with 0.5mL of 10 mM Phosphate-buffered saline (PBS) solution. That PEI solution was then added to 10 mg of UR-NPs dispersed in PBS (final volume of 2 mL). The coating was carried out at 37 ºC for 3 h under orbital stirring. The product was then washed several times with PBS, centrifuged and dried under vacuum at room temperature.

The nanoparticles here obtained were characterized by different techniques. Z potential and Dynamic Light Scattering (DLS) measurements were performed in DI water by means of a Zetasizer Nano ZS (Malvern Instruments) equipped with a 633 nm "red" laser. Fourier Transformed Infrared (FTIR) spectra were obtained in a Nicolet (Thermo Fisher Scientific) Nexus spectrometer equipped with a Smart Golden Gate ATR accessory. Transmission Electron Microscopy (TEM) was carried out with a JEOL JEM 2100 instrument operated at 200 kV, equipped with a CCD camera (KeenView Camera). Phosphotungstic acid staining was employed to detect the presence of organic matter in the obtained materials. Small Angle X-ray diffraction (SAXRD) was performed in a Philips X-Pert MPD diffractometer equipped with Cu Kα radiation.

*Cargo loading and release:* The Lower Critical Solution Temperature (LCST) behaviour of this system [20] was exploited to load a cargo in the mesopores. UR-NPs (20 mg) were placed at 4 °C and 5 mL of cargo solution (20 mg/mL fluorescein in PBS) were added and the suspension was stirred for 24 h. The sample was then filtered and washed twice with pre-heated PBS (50 °C) to remove non-loaded fluorescein. The loaded UR-NPs were then coated with PEI as described above.

For the release experiments, a suspension of 5 mg/mL fluorescein-loaded nanoparticles was prepared in 10 mM PBS pH=7.4. UR-NPs suspension was exposed to ultrasound (10 min at 1.3 MHz and 100 W in a commercial laboratory ultrasound apparatus [20]). The samples were then incubated at 37 ºC under orbital shaking for 2 h. The particles were collected by centrifugation and the amount of fluorescein released was evaluated by fluorescence spectrometry ($\lambda_{exc}$ 490, $\lambda_{em}$ 510-590 nm). Control experiments without ultrasound were performed in parallel.

**2.2 Isolation and culture of DMSCs**

Human placentas were obtained from the Department of Obstetrics and Gynecology under written informed consent approved by the Ethics Committee from *Hospital Universitario 12 de Octubre*. Processing of placental membranes and culture of primary cells was done as previously described [21]. Extra-embryonic membranes were processed by enzymatic digestion with trypsin–EDTA (Lonza, Spain). Isolated cells were grown in complete culture medium consisting of Dulbecco's modified Eagle's medium (DMEM) supplemented with 2 mM of glutamine, 0.1 mM of sodium pyruvate, 55 mM β-mercaptoethanol, 1% non-essential amino acids, 1% penicillin/ streptomycin, 10% fetal bovine serum and 10 ng/mL of EGF (epidermal growth factor), at 37 ºC, 5% $CO_2$ and 95% humidity. Non-adherent cells were discarded after 5 days. At confluence, adherent cells were passaged and seeded at a density of $10^4$ cells per $cm^2$.

**2.3 Plasmid binding to the PEI coating and transfection of DMSCs.**

*Plasmid binding:* Two plasmids were used: a plasmid encoding Green Fluorescence Protein (GFP) and a plasmid containing a suicide fusion gene consisting of the sequences for cytosine deaminase (CD) and uracil phosphoribosyl transferase (UPRT). This CD:UPRT plasmid (pSELECT-zeo-Fcy::fur) was purchased from InvivoGen (USA). For transfection, 7.5 µg of plasmid were mixed with 1 mg of UR-NPs@PEI in 1 mL of Dulbecco's Modified Eagle's Medium (DMEM), stirred at room temperature for 15 min, and then incubated in 5 mL final volume with DMSCs for 2 h at 37 ºC. The transfected cells were washed with PBS twice to remove non-internalized nanoparticles and incubated in complete culture medium.

*Transfection of GFP*: Expression of GFP was evaluated by fluorescence microscopy and flow cytometry 72 hours after transfection. Fluorescence microscopy was performed with an EVOS FL Cell Imaging System from AMG (Advance Microscopy Group). Flow cytometry was performed in a BD FACSCalibur™ cytometer (Beckton Dickinson), and data processed by Flowing Software.

*Transfection of CD:UPRT*: Expression of CD:UPRT was evaluated by DMSCs viability, reverse transcription polymerase chain reaction (RT-PCR) and 5-fluorocytosine (5-FC) conversion *in vial*.

DMSCs viability was assessed after exposure to 0.05 mg/mL of 5-FC by flow cytometry. Flow cytometry was performed using apoptosis/necrosis staining kit (BD Pharmingen™ FITC Annexin V Apoptosis Detection Kit I) following the manufacturer instructions.

In addition, expression of transfected CD:UPRT plasmid was assessed by RT-PCR. Total RNA was recovered from silica membrane columns of NZY Total RNA Isolation kit (NZYtech) after DNAse I digestion. One microgram of RNA was used for reverse transcription using High Capacity cDNA Reverse Transcription kit (Applied Biosystem). Complementary DNA (cDNA) was amplified using the kit SupremeNZY Taq (NZYtech) and primers for CD:UPRT construction (A), and human β-actin (B) as control (Table S1). PCR reactions were resolved in 2 % agarose gel and visualized by SYBR green staining.

Furthermore, expression of CD:UPRT plasmid was evaluated by 5-FC conversion *in vial*. CD enzyme activity was measured by UV-Vis spectrophotometry with a HELIOS-ZETA UV-vis spectrophotometer [15]. DMSCs ($10^5$ cells/experiment) were resuspended in 70 µL PBS and lysed by freeze-thawing. Lysates were centrifuged at 15000 rpm for 20 min at 4 ºC and 50 µL of supernatant were mixed with 50 µL of 0.4 mg/mL 5-FC in PBS and incubated at 37 ºC for 0 and 16 h. Then, 40 µL were quenched with 360 µL of 0.1M HCl and absorbance was measured at 255 and 290 nm. Enzymatic conversion of 5-FC was calculated as described [15].

**2.4 Co-culture of transfected Trojan-horse platform with NMU cancer cells.**

DMSCs transfected with CD:UPRT-loaded UR-NPs@5PEI were co-cultured with NMU rat mammary cancer cells (ATCC, LGC Standards S.L.U.) [17,18]. NMU cells were cultured in 24 well plates at a density of 20,000 cells per well. After 24 h, DMSCs (with or without nanoparticles) were seeded in Transwell® culture inserts (0.4 µm pore, polycarbonate membranes, tissue cultured treated, Costar®) in a ratio 1:2 (DMSC:NMU). After 72 hours, 5-FC was added to the medium at 0.1 mg/mL. Four days later, the inserts were removed and the viability of NMU cells was evaluated by Alamar Blue assay following the manufacturer's instructions. NMU cells were also analyzed by flow cytometry with BD Pharmingen™ FITC Annexin V Apoptosis Detection Kit I.

*Statistical analyses* were performed by the Student's *t* test.

**3. Results and Discussion**

Ultrasound-responsive mesoporous silica nanoparticles were engineered to include nucleic acids into a polycationic coating to provide gene transfection capacities. The successful synthesis of our non-viral gene transfection agent, UR-NPs@PEI, was confirmed by different characterization techniques. The prepared nanoparticles showed round-shape morphology and a diameter around 200 nm, as observed in the TEM micrographs (Figure 1a). The presence of the polymeric ultrasound-responsive gates and the successful coating with two different molecular weights of PEI was confirmed by TEM micrographs (FigureS1a, organic matter showing darker contrast by staining with phosphotungstic acid), and by FTIR spectra (Figure S1b, bands in the range 1300-1800 $cm^{-1}$). TEM micrographs also showed that the ordered porosity of the materials had been maintained throughout their preparation. Additionally, the preservation of the mesopore order (typical of MCM-41 type materials) could be observed in all prepared materials in the corresponding SAXRD patterns (Figure S1c). Successful coating of UR-NPs with both types of PEI was confirmed by the change to positive values in the zeta potential (Figure 1b). This positively-charged polymeric coating will enable the binding of negatively-charged plasmids to be used in gene transfection. Nanoparticle size measurements carried out by DLS show a hydrodynamic diameter of *ca.* 340 nm for UR-NPs@PEI, which then increased to *ca.* 530 nm after DNA complexation (Figure S2). It is worth mentioning that our nanoparticles (and their complexes with nucleic acids) will not be in direct contact with the circulatory system, since the nanoparticles will already be inside vehicle cells and non-internalized nanoparticles will be removed before injection.

To test the transfection potential of UR-NPs@1.8PEI and UR-NPs@5PEI, the GFP-expressing plasmid was used. Figure 2 shows that no significant gene expression was achieved with 1.8 kDa PEI-coated NPs. On the other hand, DMSCs transfected with UR-NPs@5PEI expressed GFP, as can be observed by fluorescence microscopy and flow cytometry. These results are in good agreement with previous literature showing that gene transfection efficiency of PEI is directly related to its molecular weight [22]. However, the toxicity of PEI is also related to its molecular weight [22], and when a 25 kDa PEI was used, the vast majority of DMSCs were substantially damaged (data not shown). Therefore, the 5 kDa PEI was chosen for all further experiments.

GFP expression was maintained in transfected DMSCs up to 14 days as was observed by fluorescence microscopy (data not shown), suggesting that gene expression would be continued after DMSCs migration to tumor [6,18].

Previous data showed that 1.8 kDa PEI coating did not affect the nanoparticle capacity of releasing a cargo upon ultrasound exposure [17]. However, we did not know whether a larger polycationic coating could hamper the correct performance of our nanodevice. To assess whether the ultrasound-responsiveness was maintained after 5 KDa PEI coating, an *in vial* cargo release experiment using fluorescein was performed (Figure 3). Fluorescence intensity in the supernatant of the samples exposed to ultrasound was highly increased compared to samples without insonation, indicating that the 5 kDa PEI coating does not prevent the opening of the polymeric gate in response to the stimulus. Upon ultrasound exposure, the monomer THPMA in the polymeric coating, p($MEO_2$MA-co-THPMA), is converted into a much more hydrophilic monomer (methacrylic acid). This change modifies the overall hydrophilicity of the system and induces a conformational change of the polymeric gate, exposing the nanoparticle pores to the environment and enabling cargo release. These results confirm the on-demand cargo release capabilities of UR-NPs@5PEI, which would enable them to carry a toxic drug preventing its premature release, which might compromise the viability of the vehicle DMSCs.

The next step after optimizing the conditions for gene transfection using UR-NPs@5PEI consisted in the use of a plasmid of therapeutic interest to be transfected in the vehicle DMSCs, to obtain our Trojan-horse platform. A plasmid containing two genes involved in suicide gene therapy, cytosine deaminase (CD) and uracil phosphoribosyltransferase (UPRT), was used. CD converts non-toxic pro-drug 5-fluorocytosine (5-FC) into toxic 5-Fluoruracil (5-FU), which is then further transformed by UPRT into the more toxic 5-Fluorouridine monophosphate (5-FUMP), an irreversible inhibitor of thymidylate synthase [23](Scheme 1).

The plasmid containing the suicide genes (CD:UPRT plasmid) was loaded into the UR-NPs@5PEI following the same protocol as previously described for the GFP plasmid. Those plasmid-loaded nanoparticles were incubated with DMSCs for 2 h and non-internalized nanoparticles were removed by washing with PBS. Seventy two hours later, 5-FC was added to DMSCs and the viability of DMSCs was evaluated after further incubation for 5 days (Figure 4). Bright field microscopy showed a reduction in cell culture density of transfected DMSCs treated with 5-FC compared to control, while no significant changes were observed with either pro-drug addition or transfection alone (Figure 4a). These results could indicate that the suicide genes are being expressed in DMSCs and the production of toxic species upon the addition of the pro-drug is hampering cell proliferation. Furthermore, apoptotic nuclei evidenced by brighter DAPI staining were observed by fluorescence microscopy in the transfected DMSCs exposed to the pro-drug (Figure 4b), as would be expected from the production of the highly toxic 5-FUMP. DMSCs exposed to those treatments were also evaluated by flow cytometry with an apoptosis-necrosis assay (Figure 4c). In each panel, cells in the lower left quadrant correspond to healthy cells, lower right quadrant are early-apoptotic cells, and upper-right quadrant comprises late-apoptotic or necrotic cells. An increase in late apoptotic/necrotic cells was observed only in the transfected DMSCs exposed to 5-FC.

Gene expression evaluated by RT-PCR confirmed the production of mRNA corresponding to the introduced sequence in DMSCs treated with a plasmid-bearing UR-NPs@5PEI (Figure S3a). The expression of the housekeeping β-actin gene was also evaluated. CD:UPRT expression was found in transfected DMSCs corresponding to three independent experiments. No CD:UPRT expression was found in control no-transfected DMSCs whereas β-actin expression was detectable in all DMSCs cultures. These data confirm the successful and highly reproducible transfection of DMSCs with CD:UPRT-bearing UR-NPs@5PEI.

The activity of the CD:UPRT enzymes in cell lysates was also assessed *in vial* (Figure S3b). A significant conversion of the pro-drug was observed with lysates from transfected DMSCs, whereas no change in 5-FC concentration was observed with control DMSCs lysates. This transformation of the pro-drug 5-FC into other species demonstrates the expression of the enzymes in their active form.

To sum up, all the above results confirm the successful transfection of DMSCs with the CD:UPRT plasmid, providing them the capability of converting non-toxic 5-FC into toxic 5-FUMP, as shown in Scheme 1. This is especially interesting because the gene transfection process has been performed with a non-viral vector under conditions where DMSCs viability has not been affected (over 90% viable DMSCs after transfection). Once the vehicle cells have reached the tumor site they will produce the toxic species upon exposure to the pro-drug, which is completely devoid of toxicity in the absence of the activating enzyme (over 94 % viable DMSCs exposed to the pro-drug) (Figure 4c).

One of the main advantages of the combination of CD and UPRT genes instead of using the CD gene alone is the increased bystander effect of 5-FUMP over 5-FU [23]. In order to test the bystander effect of our Trojan-horse platform, a Transwell co-culture with NMU mammary cancer cells was carried out [17,18]. CD:UPRT-transfected DMSCs were seeded on a Transwell cell culture insert that was placed in a well containing NMU cancer cells (1:2 ratio, DMSC:NMU). Seventy two hours after transfection of the DMSCs, the pro-drug 5-FC was added to the culture medium, and the viability of NMU cancer cells was evaluated 96 h later. Alamar Blue assay showed a significant decrease of NMU cell viability when exposed to our Trojan-horse platform in combination with 5-FC (Figure 5). These data indicate that-transfected DMSCs produce a toxic drug when exposed to non-toxic 5-FC, which diffuses from the DMSC source, reaching the surrounding cancer cells and significantly affecting their growth. Furthermore, apoptosis-necrosis analysis by flow cytometry showed a significant increase in the number of early apoptotic NMU cells when employing our Trojan-horse therapeutic strategy compared to the other control conditions (Figure 6).

In summary, the presented results confirm the successful preparation of a Trojan-horse platform capable of killing surrounding NMU cancer cells by nanoparticle-mediated genetic engineered DMSCs. The non-toxic nature of the pro-drug would ensure a good safety profile of the therapy, avoiding unnecessary exposure to a toxic molecule in healthy tissues and organs. Therefore, this anticancer therapeutic approach by the *in situ* production of a toxic molecule by tumor-tropic DMSCs transfected by non-viral nanovectors appears as a very promising tool for anticancer therapy. This effect could be additive to the previously demonstrated ultrasound-triggered release of an anticancer drug from DMSCs-transported UR-NPs [17] and the inherent anticancer effect of DMSCs [8].

## 4. Conclusions

The results obtained in this work demonstrate the possibility of inducing gene transfection using ultrasound-responsive mesoporous silica nanoparticles, without inducing significant toxicity to the vehicle cells (DMSCs), by employing a polyethylenimine coating of appropriate molecular weight. This developed non-viral transfection agent was used to transfect DMSCs with an expression plasmid containing two suicide genes, which provided them the capability of converting non-toxic 5-Fluorocytosine into toxic 5-Fluorouridine monophosphate. Moreover, DMSCs transfected with the suicide genes (Trojan-horse platform) are capable of inducing cell death in co-cultured NMU cancer cells, when exposed to the non-toxic pro-drug 5-Fluorocytosine.

## Acknowledgements

The authors thank the funding from the European Research Council through the Advanced Grant VERDI (ERC-2015 AdG proposal no. 694160). Financial support from Ministerio de Economía y Competitividad, (MEC), Spain (Project MAT2015- 64831-R) is gratefully acknowledged. This work was also funded by project PI15/01803 (Instituto de Salud Carlos III, Ministry of Economy, Industry and Competitiveness, and cofunded by the European Regional Development Fund); and by project MultimatChallenge (S2013/MIT-2862-CM, funded by the Regional Government of Madrid and EU Structural Funds), and approved by the Ethics Committee of our Institution.

**FIGURE CAPTIONS:**

**Figure 1.** TEM micrograph of UR-NPs@5PEI showing particle size and morphology as a representative example of the materials employed in this work (a), Z Potential values of UR-NPs without or with PEI coatings of two different molecular weights (1.8 or 5 kDa)(b).

**Figure 2.** Fluorescence microscopy images (left) and flow cytometry (right) of DMSCs 3 days after incubation with UR-NPs@PEI carrying the plasmid of GFP (Data are Means ± SD, N=3).

**Figure 3.** Fluorescence emission spectra of fluorescein released from UR-NPs@5PEI 2 h after ultrasound (US) application (1.3 MHz, 100 W, 10 min) (left), schematic representation of the material before and after insonation (right).

**Scheme 1.** Schematic representation of pro-drug activation. Cytosine deaminase (CD) converts 5-fluorocytosine (5-FC) into toxic 5-Fluoruracil (5-FU), which is then transformed into 5-Fluorouridine monophosphate (5-FUMP) by uracil phosphoribosyl transferase (UPRT).

**Figure 4.** Bright field microscopy images showing DMSCs after different treatments with UR-NPs@5PEI carrying the plasmid with the suicide genes CD:UPRT and/or the non-toxic pro-drug 5-FC (0.05 mg/mL) (a), DAPI-stained fluorescence microscopy images of DMSCs after the same treatments showing apoptotic nuclei (brighter nuclei staining) (b), flow cytometry analysis of apoptotis/necrosis in DMSCs after different treatments with nanoparticles carrying the suicide genes and/or the non-toxic pro-drug 5-FC (c). Scale bars represent 100 µm.

**Figure 5**. NMU cell viability results obtained by Alamar Blue assay in co-culture with DMSCs (without or with nanoparticle-mediated gene transfection) and exposed to 5-FC (Data are Means ± SD, N=3).

**Figure 6**. Apoptosis/necrosis evaluation by flow cytometry of NMU cells co-cultured with DMSCs (without or with nanoparticle-mediated gene transfection).

Figure 1

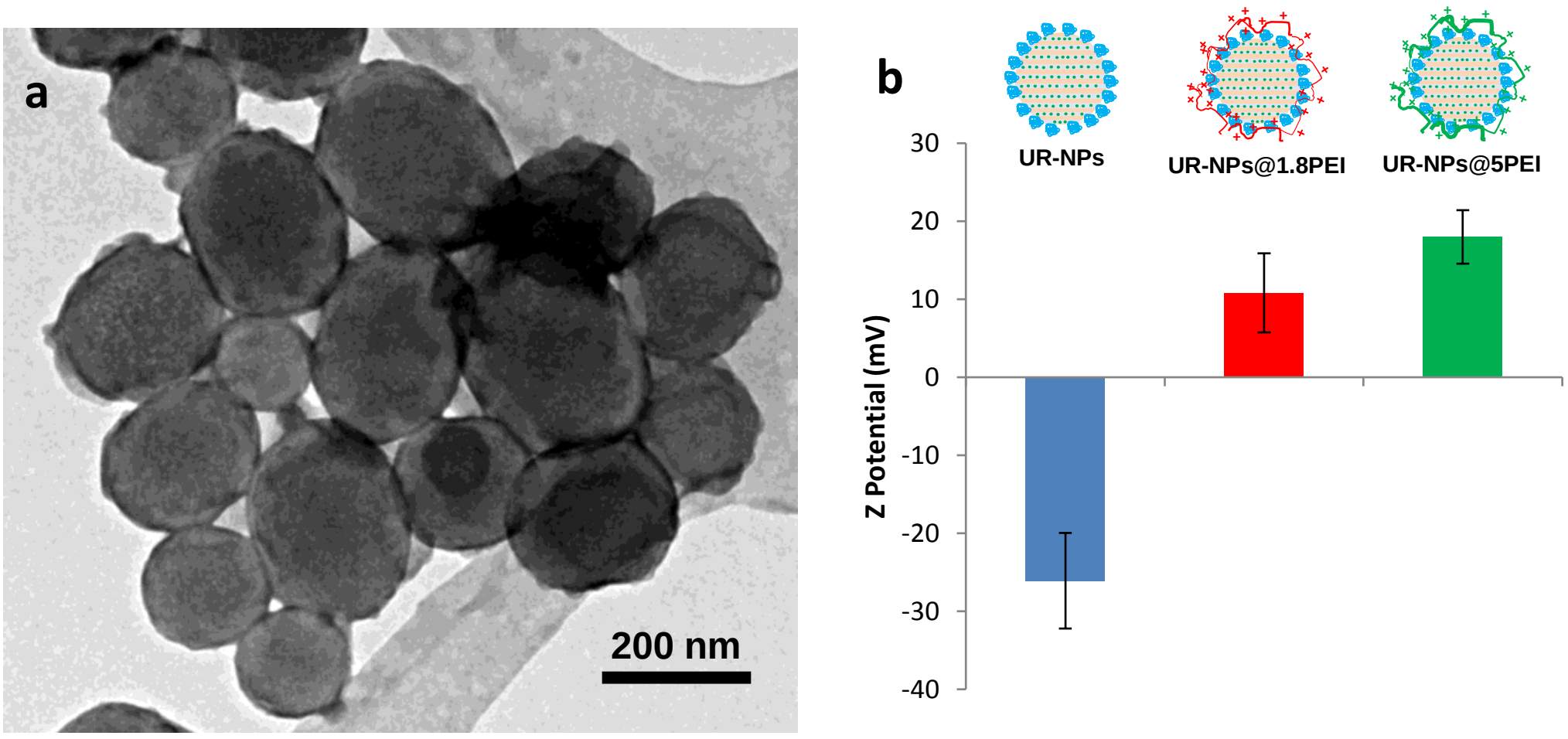

Figure 1

Figure 2

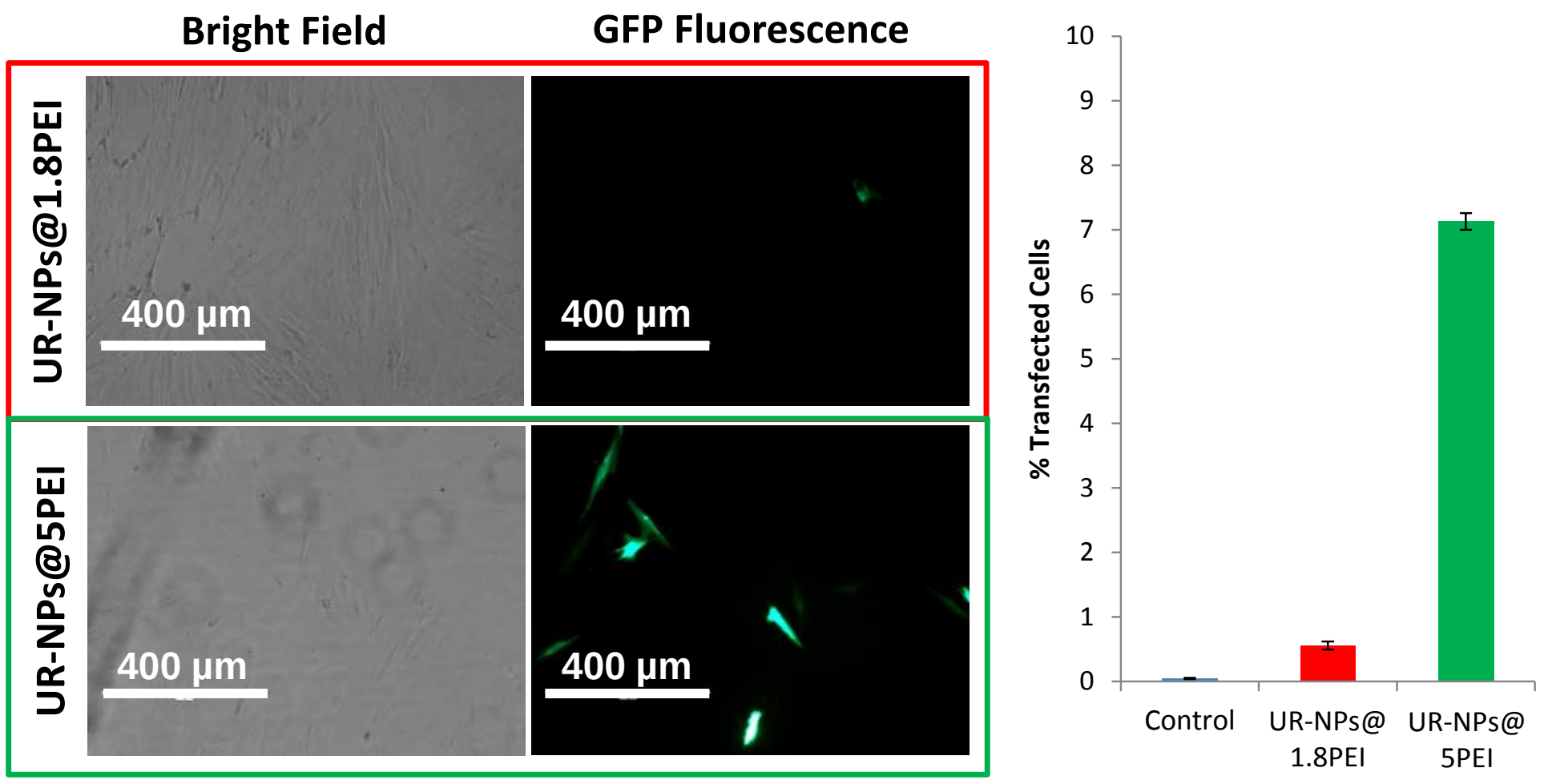

Figure 3

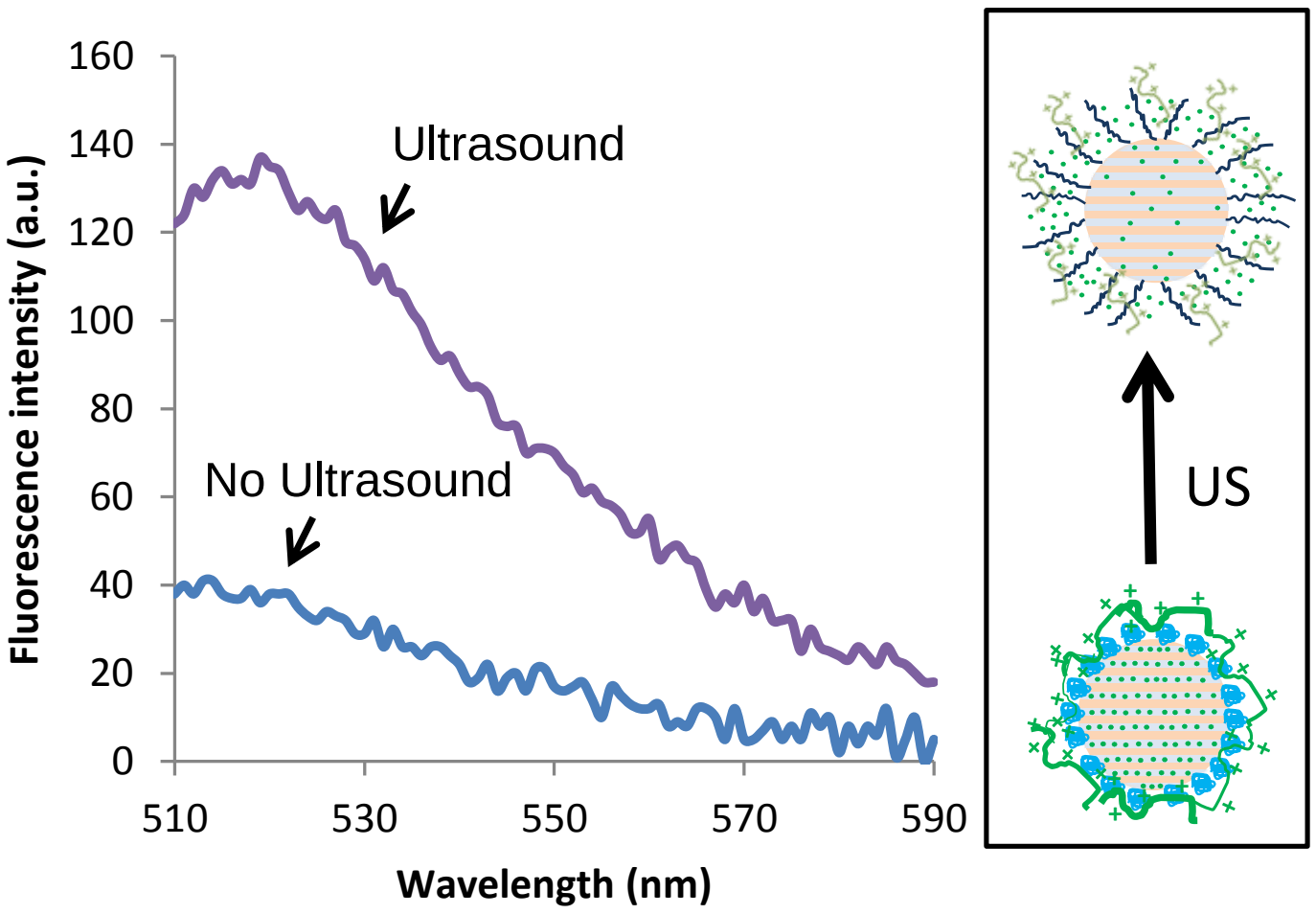

Figure 3

Figure 4

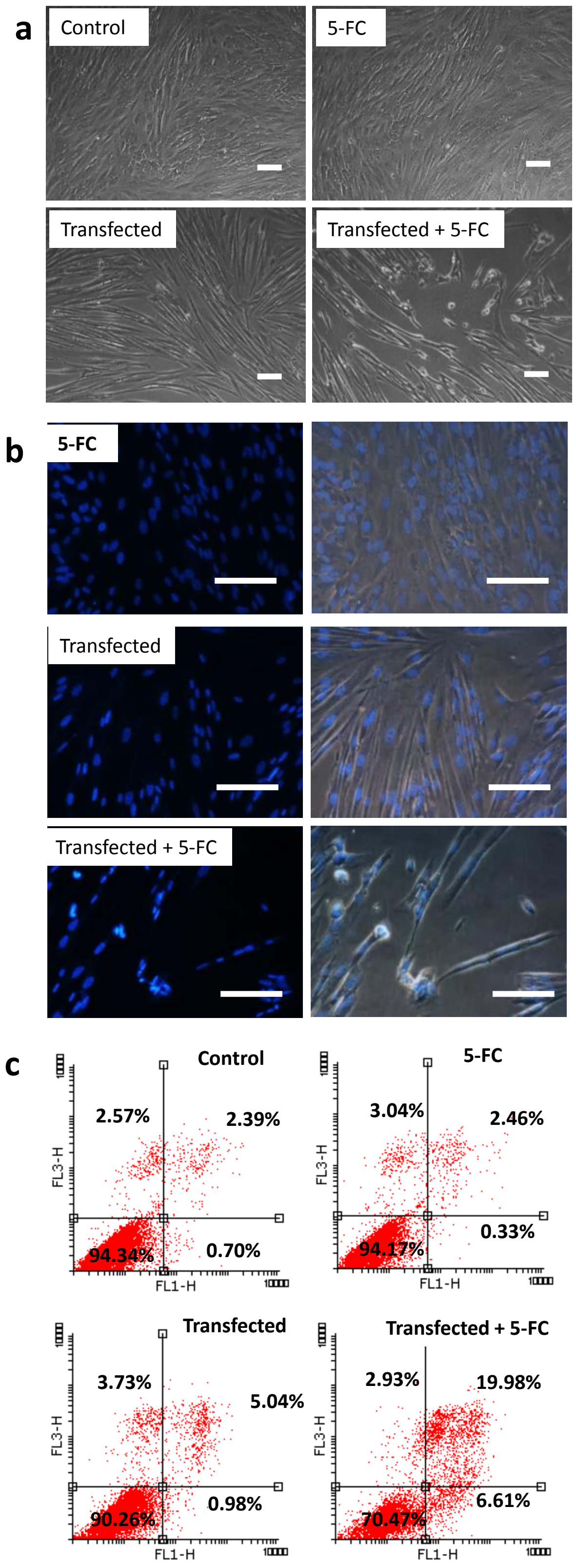

Figure 4

Figure 5

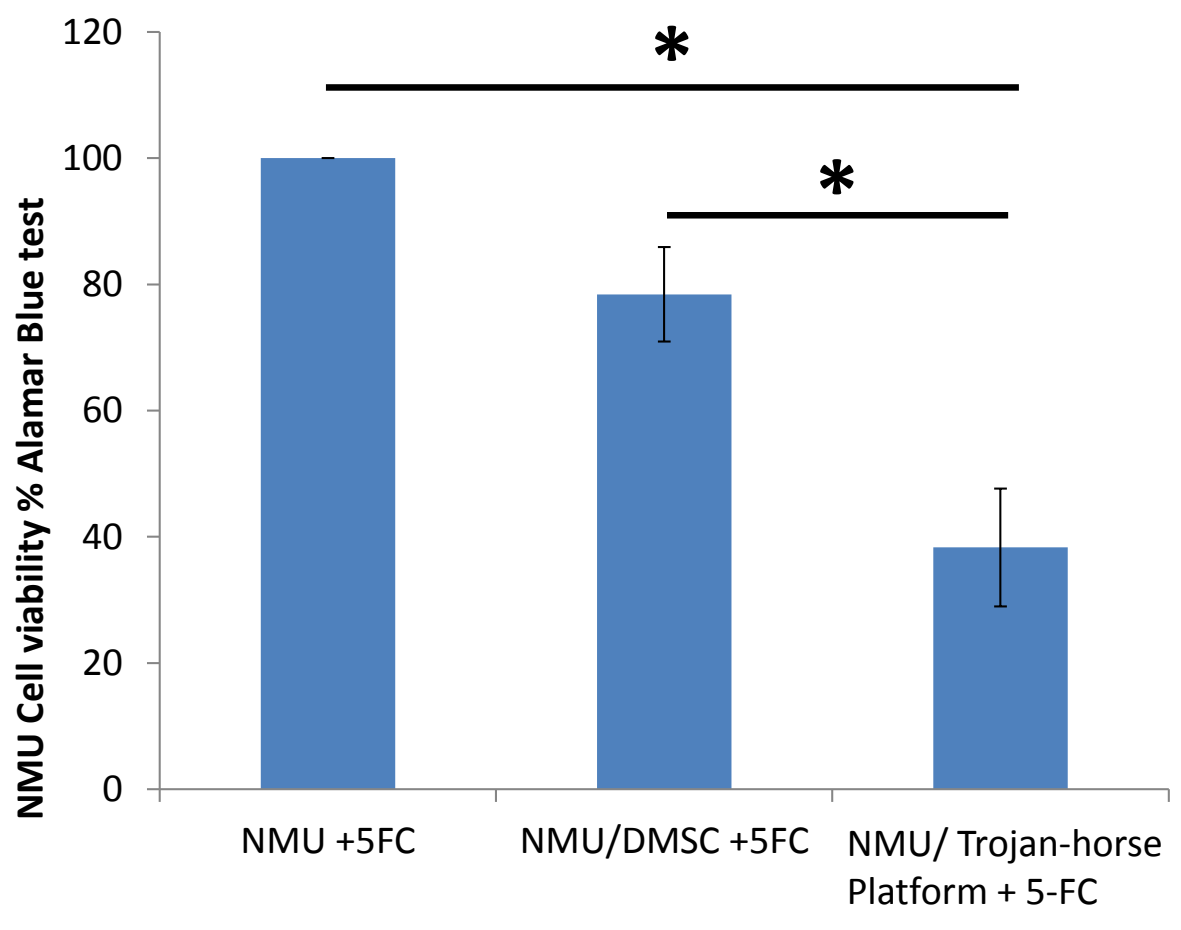

# Figure 5

Figure 6

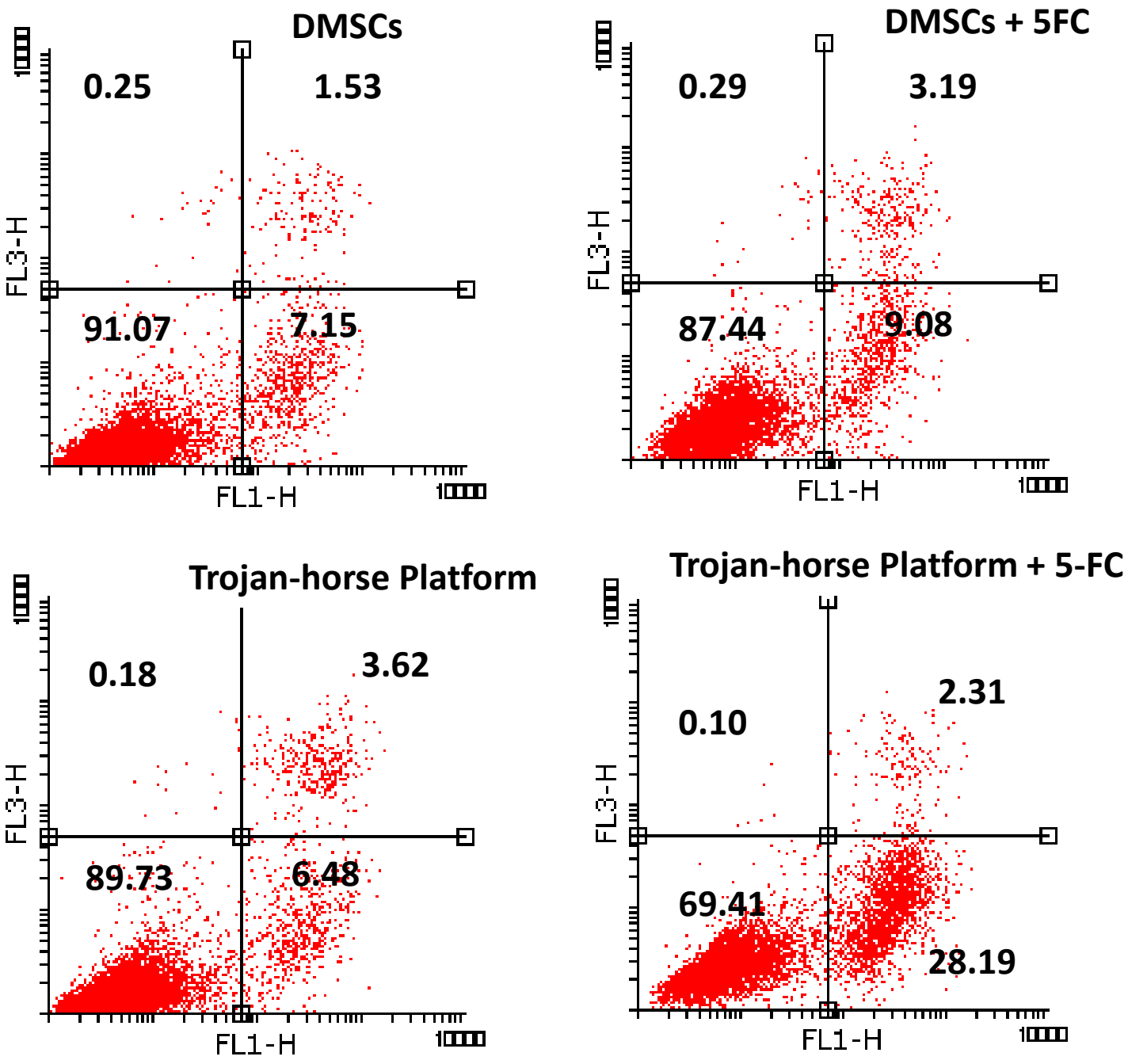

# Figure 6

Scheme 1

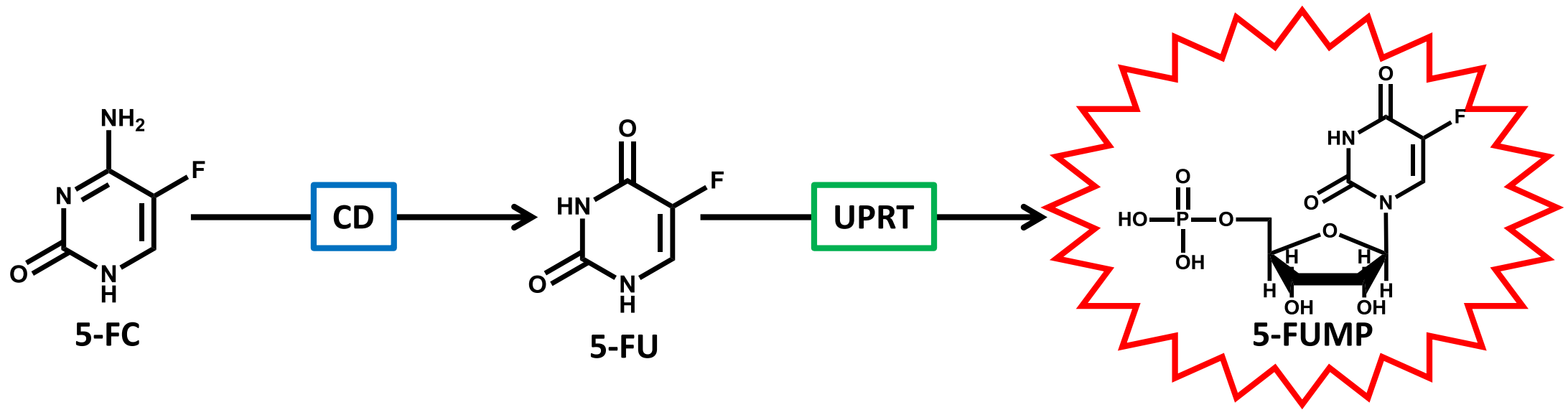